\documentclass[twocolumn]{aastex631}

\shorttitle{Vertical metallicity gradients}
\shortauthors{Long et al.}
\graphicspath{{./}{figures/}}
\begin{document}
\title{The Vertical Metallicity Gradient of the Galactic Disk for Mono-Age Stellar Populations in LAMOST}
\newcommand{\BNUFrontiers}{Institute for Frontiers in Astronomy and Astrophysics, Beijing Normal University, Beijing, 102206, China}
\newcommand{\BNUSchool}{School of Physics and Astronomy, Beijing Normal University No.19, Xinjiekouwai St., Haidian District, Beijing, 100875, China}
\newcommand{\TianjinAstroCenter}{Tianjin Astrophysics Center, Tianjin Normal University, Tianjin 300387, PR China}

\author[0009-0005-1564-0179]{Gaohuan Long}
\affiliation{\BNUFrontiers}
\affiliation{\BNUSchool}

\author[0000-0003-2471-2363]{Haibo Yuan}
\affiliation{\BNUFrontiers}
\affiliation{\BNUSchool}
\correspondingauthor{Haibo Yuan}
\email{yuanhb@bnu.edu.cn} 

\author[0000-0003-3535-504X]{Shuai Xu}
\affiliation{\BNUFrontiers}
\affiliation{\BNUSchool}

\author[0000-0002-8713-2334]{Chun Wang}
\affiliation{\TianjinAstroCenter}

\author[0000-0003-1863-1268]{Ruoyi Zhang}
\affiliation{\BNUFrontiers}
\affiliation{\BNUSchool}

\author[0000-0002-1259-0517]{Bowen Huang}
\affiliation{\BNUFrontiers}
\affiliation{\BNUSchool}

\begin{abstract}
The vertical metallicity gradient of the Galactic disk offers valuable insights into the disk's formation and chemical evolution over time. We utilized the LAMOST-LRS young stellar sample to investigate this gradient and found that it approaches zero as stellar effective temperature (or age) increases (or decreases) across various Galactocentric distances. To validate this result, we analyzed 295 open clusters younger than 3 Gyr and 976 classical cepheids within the Galactic disk. The findings confirmed that, within a given narrow age range, the vertical metallicity gradient is effectively zero. This relationship between metallicity and age supports the ``upside-down'' disk formation theory, as it indicates that the youngest and most metal-rich stars dominate the midplane, while older and more metal-poor stars formed at larger vertical heights and currently tend to be at these heights. Overall, our results align well with theoretical predictions, offering further insight into the chemical evolution and structural properties of the Milky Way.

\end{abstract}

\keywords{
    Galaxy disks (589) ;
    Galaxy structure (622) ;
    Galaxy formation (595) ;
    Galaxy evolution (594) ;
    Metallicity (1031) ;
    Galaxy abundances (574)
}

\section{Introduction} \label{sec:introduction}
Metallicity is one of the most fundamental parameters of stars and plays a crucial role in understanding the formation and evolution of galaxies \citep{Tinsley1980}. Through nucleosynthesis within stars \citep{Burbidge1957} and events such as supernova explosions \citep{Baade1934}, heavy elements are gradually injected into the interstellar medium (ISM). As stellar generations succeed one another, the metallicity of the ISM progressively increases, resulting in the formation of new stars with higher metallicities. Moreover, the distribution of stellar metallicity within the galaxies not only reflects the accumulation of heavy elements in the ISM but also provides valuable insights into the star formation history, stellar migration, and gas mixing processes in different regions of the galaxies \citep[e.g.,][]{Sellwood2002, Haywood2008}. These processes give rise to radial and vertical metallicity gradients in the Galactic disk. Therefore, studying these metallicity gradients is essential for reconstructing the chemical evolution and dynamical history of the Milky Way (MW).
 
For the radial metallicity gradient in the Galactic disk, extensive studies have been conducted using various stellar samples. Numerous studies have reported different findings in various stellar populations, although a common conclusion is the presence of a negative gradient (e.g., \citealt{Chen2003}, \citealt{Daflon2004}, \citealt{Maciel2009}, \citealt{Bilir2012}, \citealt{Coskunoglu2012}, \citealt{Boeche2013}, \citealt{Huang2015}, \citealt{OnalTas2016}). Some studies have identified variations in gradient steepness according to population age, with younger populations generally exhibiting steeper radial gradients \citep[e.g., ][]{Xiang2015, Vickers2021, Lian2022, Lian2023}. 

In contrast, studies on the vertical metallicity gradient in the Galactic disk are relatively scarce. Existing research reveals that the vertical metallicity gradient in the Galactic disk is typically negative (e.g., \citealt{Karaali2003}, \citealt{Ak2007}, \citealt{Ak2007b}, \citealt{Chen2011}, \citealt{Kordopatis2011}, \citealt{Schlesinger2012}, \citealt{Schlesinger2014}, \citealt{Imig2023}). However, despite revealing this general trend and showing that metallicity is closely related to stellar age, these studies face limitations due to their reliance on mixed-age stellar samples, which makes it difficult to accurately trace the evolution of vertical gradients over time \citep[e.g.,][]{Chen2011, Hayden2014}.

Observational studies by \cite{Xiang2015} indicate that the vertical metallicity gradient derived from a sample that includes stars of all ages is significantly steeper than that obtained from stars in individual age intervals. Their analysis, based on relatively broad age intervals that encompass multiple stellar populations, may obscure the true vertical metallicity distribution for stars of different ages. Specifically, younger stars are typically located closer to the Galactic disk with higher metallicities, whereas older stars are found farther from the plane with lower metallicities, meaning that age mixing can cause observed gradients to appear artificially steep. 

Numerical simulations by \cite{Graf2024} show that the vertical metallicity gradient of mono-age stellar populations is significantly smaller than that of mixed-age stellar samples, which aligns with the observational trend. These simulations further suggest that the vertical metallicity gradient in the Galactic disk primarily reflects an ``upside-down'' \citep{Bird2013} settling process over time.

However, the classification of mono-age stellar populations often involves significant age uncertainties, which can introduce biases due to age mixing when investigating the temporal evolution of vertical metallicity gradient. Compared to older stars, younger stars are less affected by radial migration and long-term dynamical evolution \citep[e.g.,][]{Minchev2010, Kubryk2015, Frankel2018, Frankel2020, Lian2022}. Additionally, high-temperature young main-sequence stars exhibit a stronger correlation between effective temperature and age compared to low-temperature stars \citep[e.g.,][]{Schaller1992, Zorec2012, Sun2021, Wang2023}, providing theoretical support for using temperature as a basis for stellar grouping.

In this study, we use low-resolution spectral (LRS) data from the Large Sky Area Multi-Object Fibre Spectroscopic Telescope (LAMOST; \citealt{Cui2012}) to construct four mono-temperature stellar populations and investigate their vertical metallicity gradients. By dividing stars into $T_{\mathrm{eff}}$ intervals, our approach effectively reduces the bias introduced by age mixing, yielding vertical metallicity gradients that align more closely with numerical simulations, thereby offering a more precise characterization of the chemical evolution in the Galactic disk. Compared with previous observational studies, our research reveals a clearer pattern in the temporal evolution of vertical metallicity gradients based on a more refined age classification.

The paper is organized as follows. Section 2 describes the data used in our study. Section 3 outlines the methods employed. In Section 4, we present and validate the results, supplemented by data from open clusters and classical cepheids. Finally, Section 5 provides a summary and discusses future perspectives.

\section{Data} 
\subsection{Data Selection and Quality Control}
In our study of the vertical gradient of [Fe/H] in the Galactic disk, young main-sequence stars are ideal targets. Therefore, we adopt the results from \cite{Xiang2022} as our main sample, who used the Payne method to estimate the parameters of OBA-type hot stars (OBA stars) in the sixth data release (DR6) of the LAMOST. This method provides 11 stellar parameters, including $T_{\mathrm{eff}}$, $\log g$, and [Fe/H], for approximately 330,000 hot stars. 

For a comparative study, we also select a sample of cooler main-sequence G-type dwarfs (G dwarfs) with temperatures primarily in the $5500 < T_{\text{eff}} < 6000 \, \mathrm{K}$ range \citep{Eker2018, Eker2020} from the ninth data release (DR9) of the LAMOST. This temperature range is characterized by significant age mixing, which impacts the resulting vertical metallicity gradients. By including these G dwarfs, we aim to examine how age mixing, rather than merely differences between young and old stars, influences the observed vertical gradients. This sample serves as a comparison to the younger, more homogeneous populations analyzed in the LAMOST-LRS dataset, helping us better understand the role of age distribution in shaping the vertical metallicity gradient. In this study, we cross-match the OBA stars and G dwarfs with Gaia DR3 \citep{GaiaCollaboration2023} to obtain their proper motion components (pmra and pmdec). 

To ensure data quality, we apply the following selection criteria to the OBA star catalog:
\begin{enumerate}
    \item Signal-to-noise ratio: (S/N)$_g$ $>$ 30.
    \item chi2ratio $<$ 1, as recommended by \citet{Xiang2022}.
    \item $\sigma$[Fe/H] $<$ 0.2\,dex to limit uncertainties in metallicity.
    \item Remove chemically peculiar stars by excluding those with $\mathrm{[Fe/H]} > -1.1 \times 10^{-4} \times T_{\mathrm{eff}} + 1.4$.
    \item $7000 < T_{\mathrm{eff}} < 15,000\, \mathrm{K}$, $3.5 < \log g < 5$, and $-1.2 <$ [Fe/H] $< 0.5$.
    \item Vertical distance from the Galactic plane: $|Z| < 1$ kpc.
    \item To mitigate the potential impact of unresolved binary systems, only stars with \texttt{RUWE} $<$ 1.4 from Gaia DR3 are included in the sample \citep{CastroGinard2024}.
\end{enumerate}
As a result, we obtain a final sample of 129,245 LAMOST OBA stars.

\begin{figure*}[htbp]
   \centering
   \includegraphics[width=\linewidth]{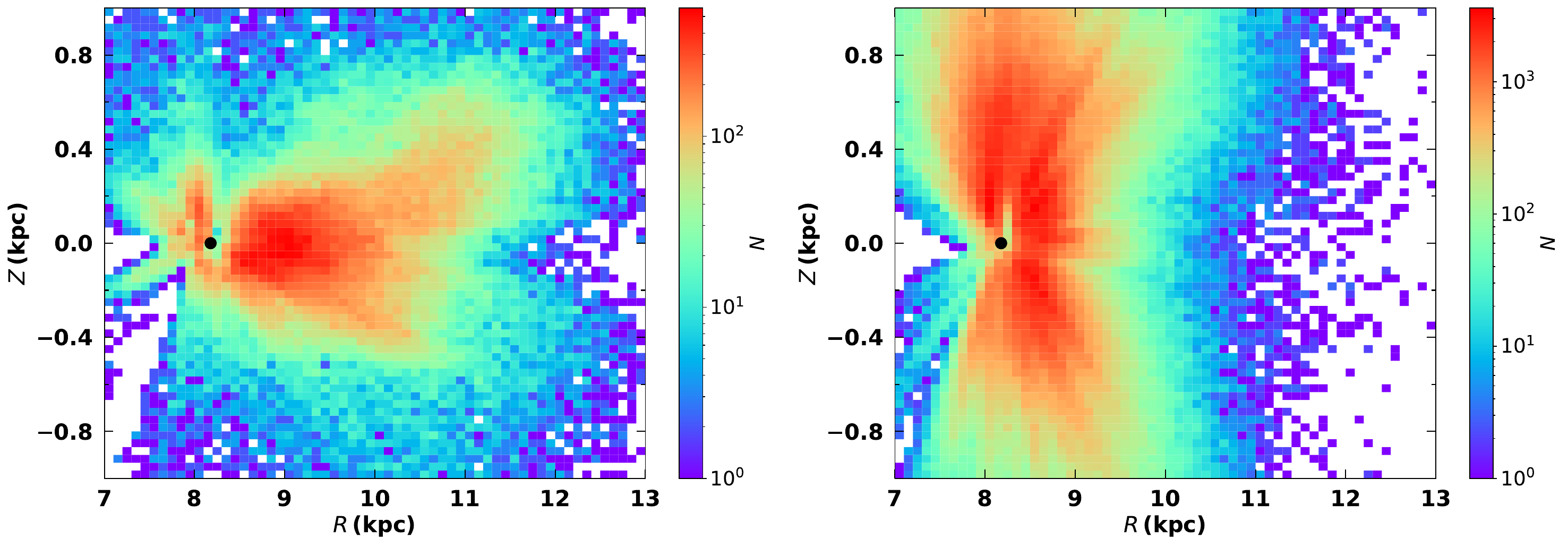} 
   \caption{The spatial distributions in the $R$-$Z$ plane of the selected OBA stars (left) and G dwarfs (right). The Sun, marked by a black dot, is located at $(R, Z) = (8.178, 0)$ kpc. The colors indicate number densities, as shown in the color bars.
}
   \label{fig:1}
\end{figure*}

For the LAMOST DR9, we apply the selection criteria $5500 < T_{\text{eff}} < 6000 \, \mathrm{K}$ and $\log g > 4.1$ and RUWE $<$ 1.4 from Gaia DR3, resulting in a sample of 954,760 G dwarfs. The stellar number density distribution on the $R$–$Z$ plane is shown in Figure~\ref{fig:1}.

\subsection{Correcting for the Temperature-Dependent Systematic Errors of [Fe/H]}
\begin{figure*}[htbp]
   \centering
   \includegraphics[width=\linewidth]{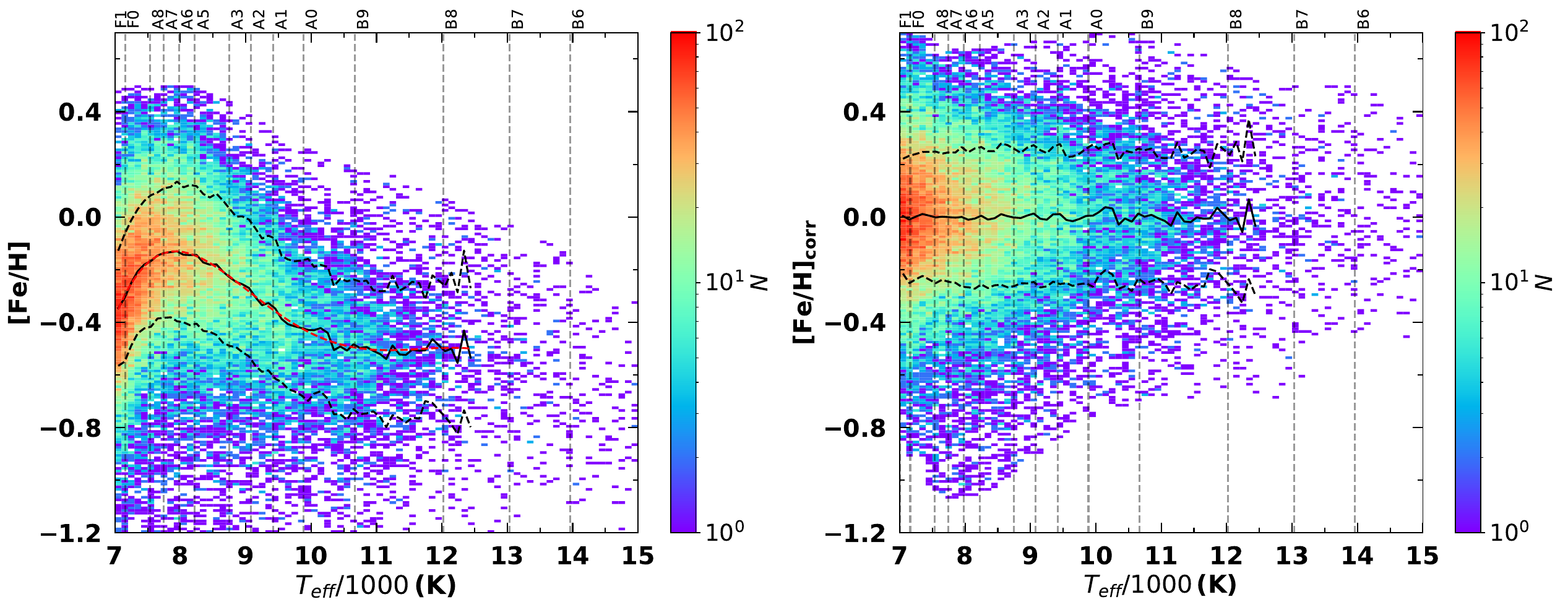} 
   \caption{The temperature-dependent systematic errors of [Fe/H] for the OBA stars before (left panel) and after (right panel) correction. The colors indicate number densities, as shown in the color bars. The black lines represent the median (solid line) and standard deviations (dashed lines), while the red line in the left panel illustrates the polynomial fit result. The vertical dashed lines indicate the temperature ranges of OBA stars, with corresponding spectral classes labeled \citep{Eker2018}.
   }
   \label{fig:2}
\end{figure*}

To address potential temperature-dependent systematic biases in [Fe/H] measurements for OBA stars, we adopt the empirical correction scheme proposed by \citet{Wang2023} through the following procedure‌. To minimize the influence from the Galactic radial metallicity gradient, we exclusively select stellar targets within the galactocentric distance range of $8.7 < R < 9.5$ kpc, where no significant physical trend between $T_{\mathrm{eff}}$ and [Fe/H] is expected for the OBA stars. We then implement a sixth-order polynomial to model the observed trend between $T_{\mathrm{eff}}$ and [Fe/H], which is primarily attributed to non-local thermodynamic equilibrium (NLTE) effects‌. As visually confirmed in Figure~\ref{fig:2}, the systematic bias displays a pronounced temperature dependence prior to correction, while the trend is flat after correction. We subsequently extend this calibrated correction to our complete stellar sample, ensuring metallicity measurements remain consistent across different temperature ranges. The metallicity used in the following analysis is the corrected metallicity ([Fe/H]$_{\text{corr}}$).

For the G dwarfs, we apply the calibration curve proposed by \citet{Niu2023} (see their Section 3.2) to correct for the temperature-dependent biases in [Fe/H] measurements from the LAMOST LRS.

\section{Method} \label{sec:floats}
\subsection{Coordinate Systems and Galactic Parameters}
In this study, we adopt a Galactocentric cylindrical coordinate system $(R, \phi, Z)$, where $R$ represents the projected Galactocentric distance, $\phi$ denotes the azimuthal angle between the line from the Galactic center to the Sun and the line to the projected stellar position, and $Z$ indicates the height above or below the true Galactic plane, with positive values pointing toward the North Galactic Pole. We use the photogeometric distances provided by \cite{Bailer-Jones2021} to derive the spatial positions in this coordinate system. In our analysis, we assume the Sun resides on the Galactic midplane ($Z_\odot = 0$ pc) at a Galactocentric distance of $R_\odot = 8.178$ kpc \citep{GRAVITY2019}.

\subsection{The \texorpdfstring{$Z_{\text{max}}$}{Zmax} Parameter}
We investigate one of the key parameters in stellar orbital dynamics, $Z_{\text{max}}$, which represents the maximum vertical distance a star can reach relative to the Galactic plane during its orbit. $Z_{\text{max}}$ reflects both the initial conditions of a star and the cumulative effects of its dynamical evolution. We utilize the Action-based Galaxy Modelling Architecture (AGAMA) package \citep{Vasiliev2019} to compute $Z_{\text{max}}$. To ensure reliable orbital parameters, we integrate stellar orbits over a time span of 10 Gyr, allowing most stars to complete at least one full revolution around the Galactic center. The integration is performed with 101 time steps (\texttt{numsteps}=101) to maintain numerical stability and convergence. In our calculations, we adopt the Galactic potential model from \cite{McMillan2017}, which includes contributions from the disk, bulge, and halo, providing a well-validated representation of the MW’s gravitational field \citep[e.g.,][]{Yuan2024, Bonaca2025}.

\subsection{Subtracting the Radial Metallicity Gradients}
To minimize the impact of age-mixing effects on the measurement of the vertical gradient of [Fe/H], we divide the OBA stars into four stellar groups based on $T_{\text{eff}}$, with ranges of $7000 < T_{\text{eff}} < 8000 \, \mathrm{K}$, $8000 < T_{\text{eff}} < 9000 \, \mathrm{K}$, $9000 < T_{\text{eff}} < 10{,}500 \, \mathrm{K}$, and $10{,}500 < T_{\text{eff}} < 15{,}000 \, \mathrm{K}$ \citep{Eker2018, Eker2020}. As the $T_{\mathrm{eff}}$ range increases, the age span within each interval becomes narrower, progressively approaching the characteristics of strictly mono-age stellar populations. The presence of the radial gradient of [Fe/H] may interfere with the precise measurement of the vertical gradient. To eliminate such effect, we subtract the trend of radial gradient of our samples. 

\begin{figure*}[htbp]
   \centering
   \includegraphics[width=\linewidth]{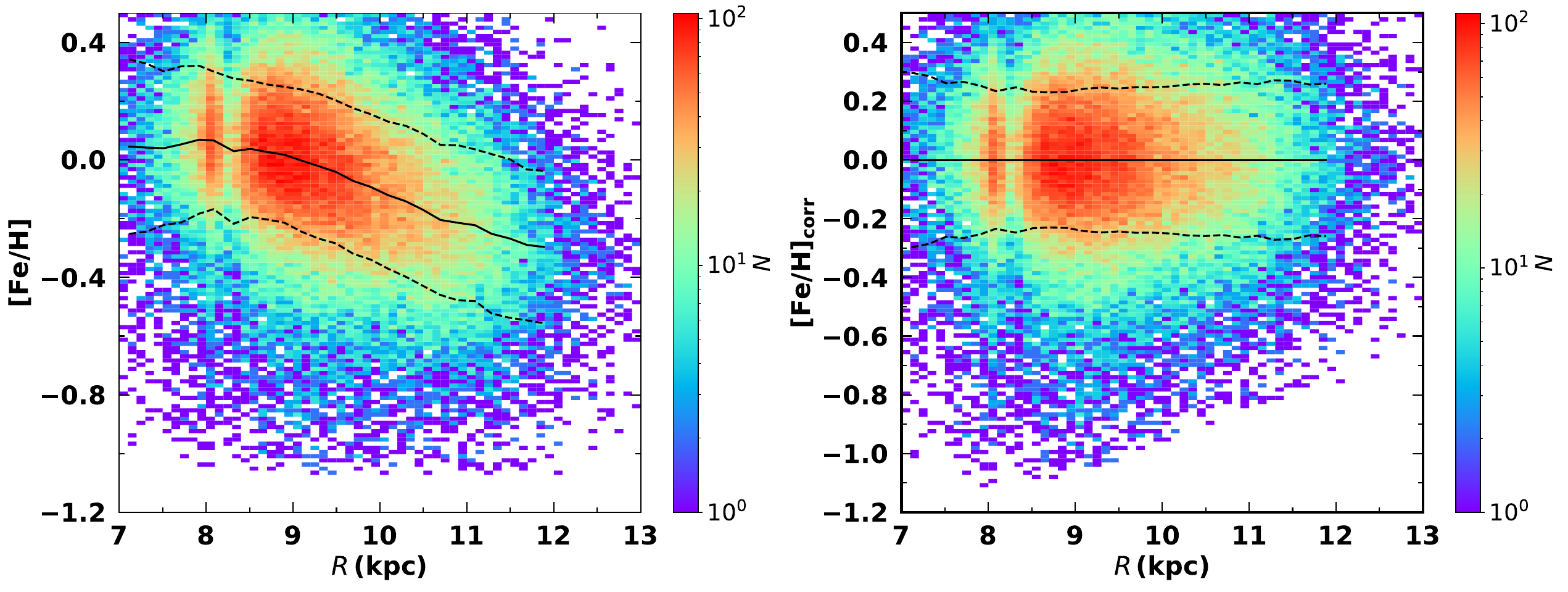} 
   \caption{The variation of [Fe/H] with $R$ for the OBA stars in the range $7000 \, \mathrm{K} < T_{\mathrm{eff}} < 8000 \, \mathrm{K}$ before (left panel) and after (right panel) correction. The colors indicate number densities, as shown in the color bars. The black lines represent the median (solid line) and standard deviations (dashed lines).}
   \label{fig:3}
\end{figure*}

For the G dwarfs, we select only stars with $|Z|$ less than 0.2 kpc and then perform a linear fit of metallicity with $R$ as the trend for the whole sample. For the OBA stars,  we subtract the median trend of metallicity with $R$ separately for different temperature intervals. An example is shown in Figure~\ref{fig:3}. The metallicity values used in this study refer to those after radial trend correction.

\section{Result}\label{sec:displaymath}
\subsection{Vertical Gradients of [Fe/H]}
In this section, we analyze the vertical gradient of [Fe/H] for mono-temperature stellar populations. We apply the aforementioned temperature ranges to divide the OBA stars into different populations. The analysis is conducted within three ranges of $R$: $7 < R \leq 9$ kpc, $9 < R \leq 11$ kpc, $11 < R \leq 13$ kpc.

\begin{figure*}[htbp]
   \centering
   \includegraphics[width=\linewidth]{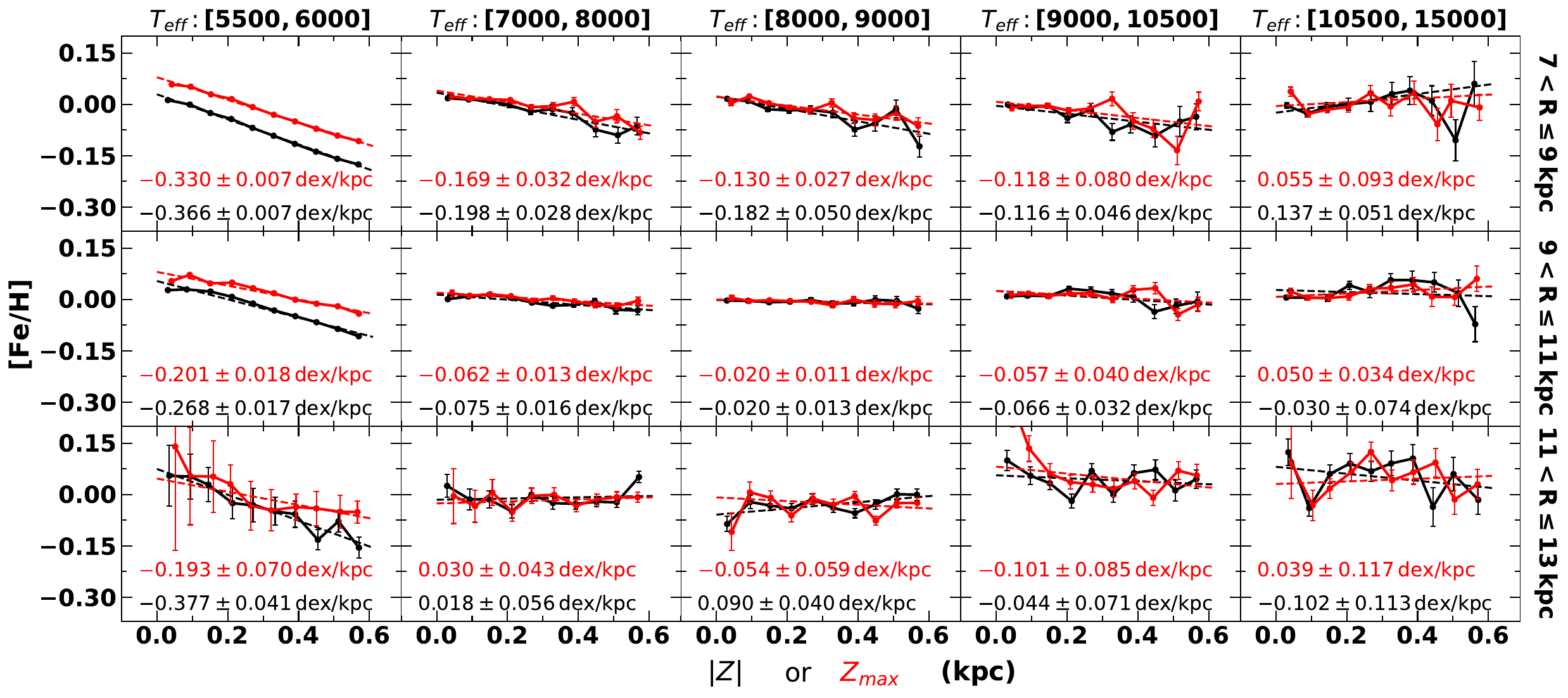} 
   \caption{The median metallicity values  as functions of $|Z|$ (black lines) and $Z_{\text{max}}$ (red lines) for the five mono-temperature stellar populations and at different $R$ ranges. The black and red dashed lines represent the linear fit results with $|Z|$ and $Z_{\text{max}}$, respectively. The slopes and corresponding errors are labelled.}
   \label{fig:4}
\end{figure*}

Figure~\ref{fig:4} shows the trends of the median [Fe/H] as a function of $|Z|$ (black lines) and $Z_{\text{max}}$ (red lines) for five mono-temperature stellar populations across different $R$ ranges. Each point represents the median [Fe/H] calculated within a 0.06 kpc interval. The dashed lines represent the linear regression results of these medians. The trends of metallicity with $Z_{\text{max}}$ and $|Z|$ are generally consistent, but the former usually exhibits slightly shallower gradients and better stability.

The G dwarfs exhibit a significant vertical metallicity gradient due to their broader age range. This steep gradient is primarily determined by two factors: Firstly, according to the ``upside-down'' formation theory, older stars generally form in dynamically active, gas-rich, and metal-poor environments, leading to lower metallicities at higher $Z_{\text{max}}$ and $|Z|$ positions \citep{Bird2013}. Secondly, early Galactic dynamical heating events, such as galaxy mergers and gravitational perturbations, have disturbed stellar orbits, pushing stars from the original thin disk to higher orbits, thereby creating a thicker and more dispersed stellar distribution than the original disk \citep{Graf2024}. Moreover, the continuous formation and evolution of multiple generations of stars within the Galaxy means that G dwarfs cover a wide range of metallicities. 

In contrast, OBA stars show a progressively weaker vertical metallicity gradient as their temperature increases (and age distribution narrows), with the gradient approaching zero, particularly in the hottest stellar populations. This trend is consistent with the theoretical expectation that in a strict mono-age stellar population, the vertical metallicity gradient would be zero.

Furthermore, for a given temperature range, the gradient is relatively steep in the smaller $R$ range ($7 < R \leq 9\,\text{kpc}$), while it flattens as $R$ increases. This radial dependence can be attributed to the systematic variation in stellar age distributions: regions with smaller radii include a wider age range, thereby encompassing a greater portion of the Galactic disk's evolutionary history. In contrast, regions with larger radii have narrower age distributions, predominantly consisting of young stars, which leads to a significant weakening of the metallicity gradient. This radial dependence reflects not only the ``upside-down'' but also the "inside-out" nature of disk formation. 

\begin{figure*}[htbp]
   \centering
   \includegraphics[width=\linewidth]{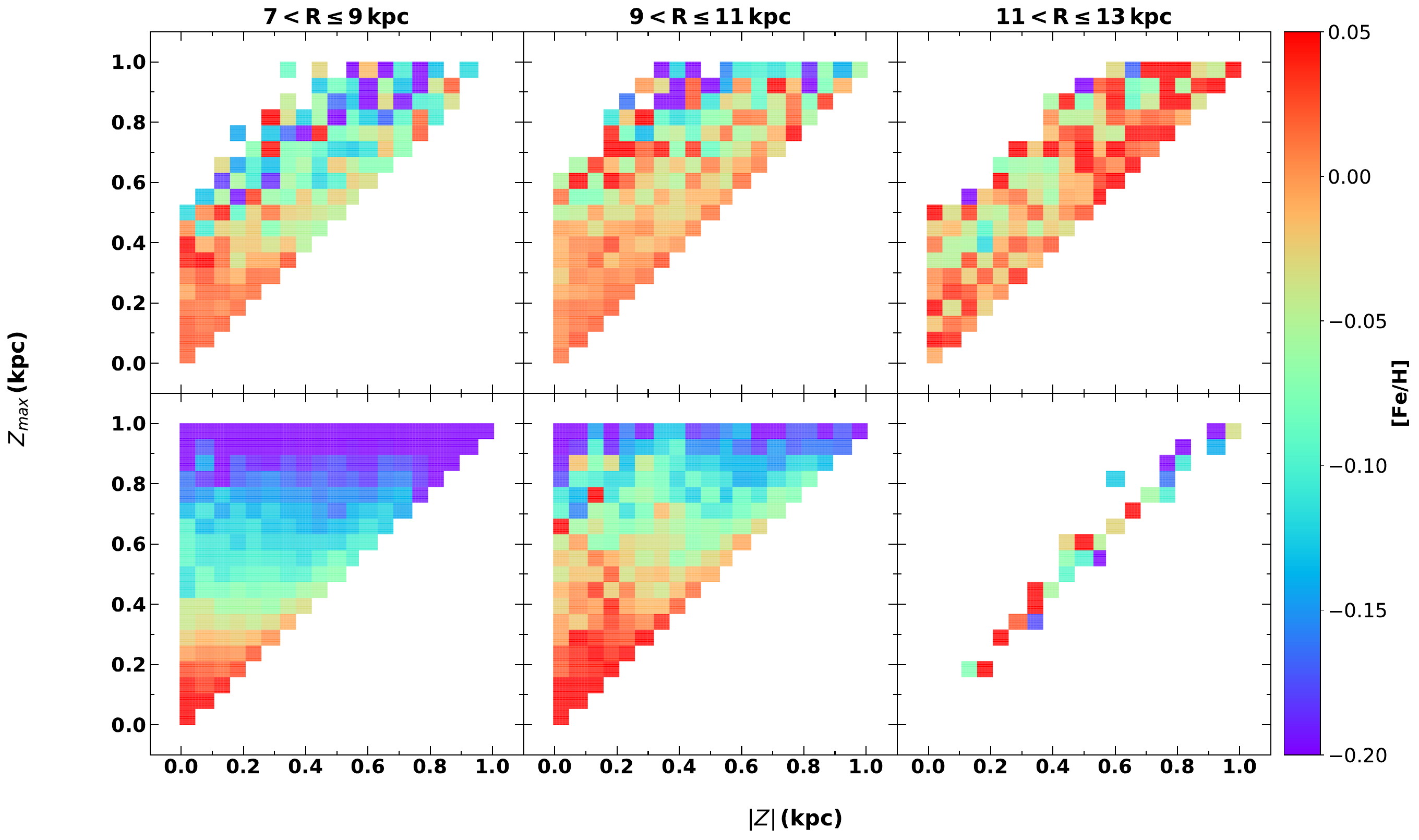} 
   \caption{The [Fe/H] distributions in the $|Z|-Z_{\text{max}}$ plane of the mono-temperature stellar populations are shown at different $R$: $7 < R \leq 9\,\text{kpc}$, $9 < R \leq 11\,\text{kpc}$, and $11 < R \leq 13\,\text{kpc}$. The top and bottom panels represent the OBA stars and the G dwarfs, respectively. The colors indicate [Fe/H], as shown in the color bars.}
   \label{fig:5}
\end{figure*}

Figure~\ref{fig:5} presents the [Fe/H] distribution of OBA stars (top panels) and G dwarfs (bottom panels) in the $|Z|-Z_{\text{max}}$ plane, divided into three radial bins: $7 < R \leq 9\,\text{kpc}$, $9 < R \leq 11\,\text{kpc}$, and $11 < R \leq 13\,\text{kpc}$. For both OBA stars and G dwarfs, within a given narrow range of $Z_{\text{max}}$, the [Fe/H] distribution exhibits weak variations with $|Z|$. The result further suggests that the vertical metallicity gradient of mono-age stellar populations is consistent with zero.

In addition, we have assessed the influence of the Galactic warp \citep{Chen2019} on our result. Since our sample is primarily located in the anti-Galactic center direction where the warp is weak, the warp's influence on the vertical metallicity gradient is negligible.

\subsection{Supplementary Validation Using Open Clusters and Classical Cepheids}
Open clusters (OCs) are ideal objects for investigating the vertical gradient of [Fe/H] in the Galactic disk due to their relatively young ages and precise age estimates. To achieve this, we use high-quality data from \cite{Zhong2020}, obtain by cross-matching the LAMOST DR5 spectroscopic catalog with the OCs identified by \cite{Cantat-Gaudin2018}. 

\begin{figure}[htbp]
   \centering  \includegraphics[width=1\linewidth]{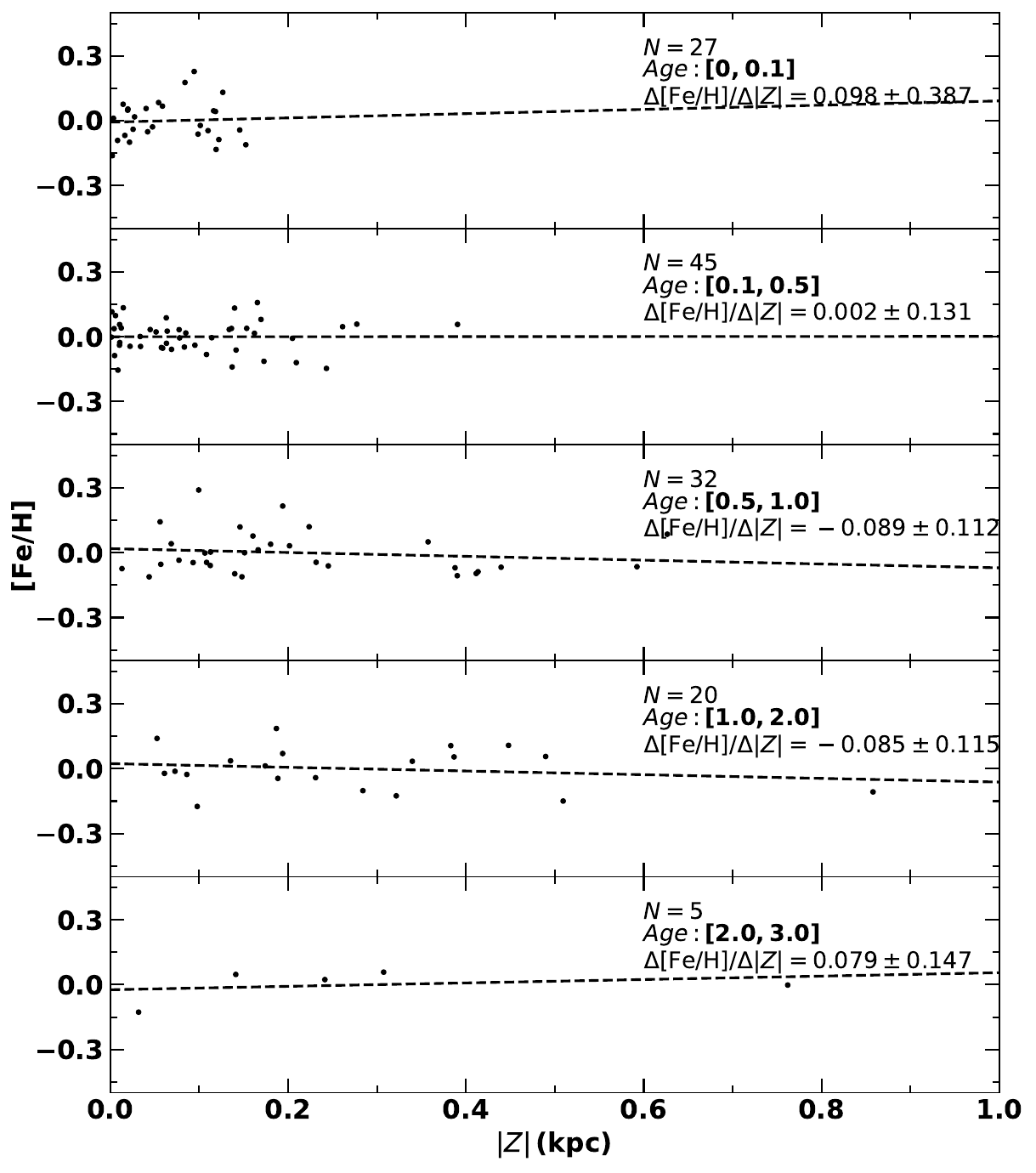} 
   \caption{Vertical metallicity gradients of OCs in different age bins. The dashed lines represent linear fitting results, with the corresponding slopes labeled.}
   \label{fig:6}
\end{figure}

To remove the vertical [Fe/H] gradient from the radial influences, a key improvement over \citet{Zhong2020} - we first remove the radial [Fe/H] gradient and then divide the OCs into five narrow age groups: $\text{age} < 0.1\,\text{Gyr}$, $0.1 < \text{age} < 0.5\,\text{Gyr}$, $0.5 < \text{age} < 1.0\,\text{Gyr}$, $1.0 < \text{age} < 2.0\,\text{Gyr}$, and $2.0 < \text{age} < 3.0\,\text{Gyr}$. Figure~\ref{fig:6} shows the vertical gradient of [Fe/H] for each age group. The slope varies between $-0.089$ and $0.098 \, \text{dex} \, \text{kpc}^{-1}$, consistent with zero within error bars for each age group. 

Compared to the results of \citet{Zhong2020}, our study shows that the vertical gradient of [Fe/H] of mono-age stellar populations is nearly zero within the error range. For example, for OCs between \(0.1 < \text{age} < 0.5 \, \text{Gyr}\) and \(0.5 < \text{age} < 1 \, \text{Gyr}\), \citet{Zhong2020} report gradients of \(-0.122 \pm 0.150 \, \text{dex} \, \text{kpc}^{-1}\) and \(-0.338 \pm 0.086 \, \text{dex} \, \text{kpc}^{-1}\), respectively. In contrast, after taking into account the influence of the radial gradient of [Fe/H], we measure gradients of \(0.002 \pm 0.131 \, \text{dex} \, \text{kpc}^{-1}\) and \(-0.089 \pm 0.112 \, \text{dex} \, \text{kpc}^{-1}\). This indicates that, after accounting for the radial gradient, our results are closer to the near-zero vertical metallicity gradient observed for mono-temperature stellar populations. This finding not only validates the necessity of removing the radial gradient in our analysis but also emphasizes the advantages of our method, laying a foundation for applying this approach to study the chemical evolution of the MW in larger samples.

\begin{figure}[htbp]
   \centering   \includegraphics[width=1\linewidth]{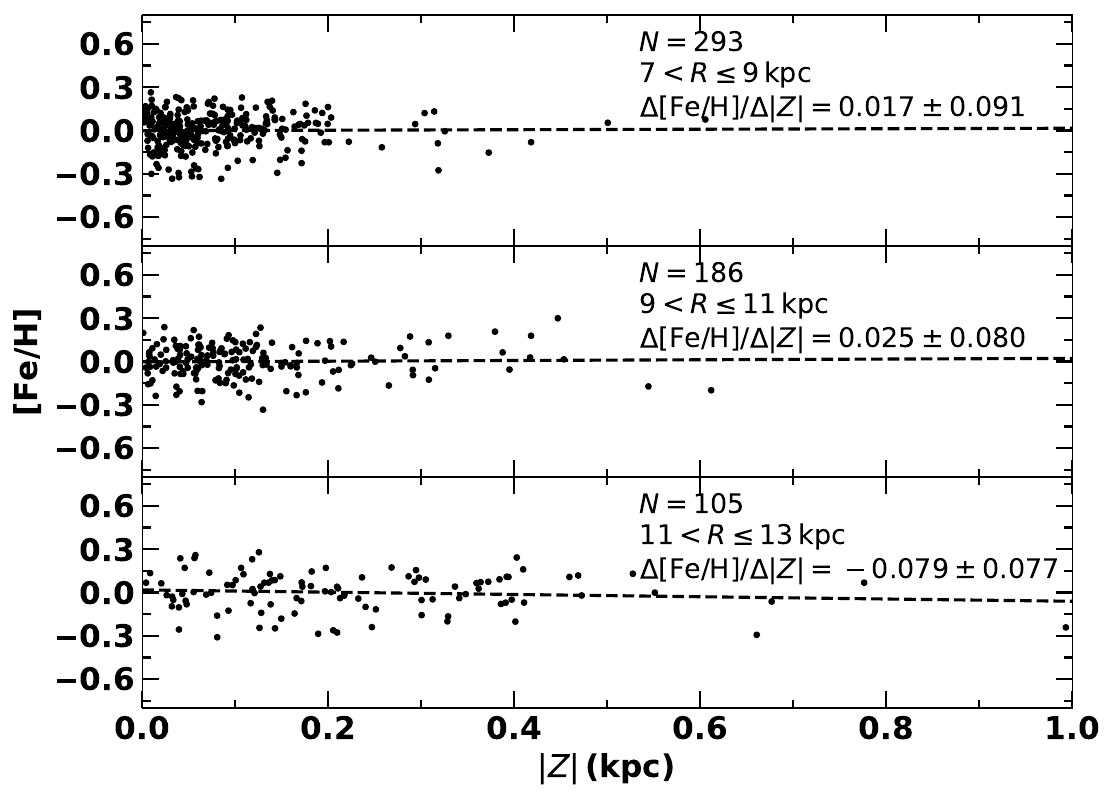} 
   \caption{Vertical metallicity gradients in different $R$ bins for DCEPs. The dashed lines represent linear fitting results, with the corresponding slopes labeled.}
   \label{fig:7}
\end{figure}

Classical cepheids (DCEPs) are among the youngest stars in the Galaxy. To supplement our study, we utilize the cepheid catalog provided by Gaia DR3 \citep{GaiaCollaboration2023}. This catalog contains the spectroscopic [Fe/H] values and their uncertainties for 949 DCEPs in the Gaia DR3 sample, as well as for 27 DCEPs from the literature. We select sources with \( |Z| < 1 \, \text{kpc} \), $7 < R \leq 13\,\text{kpc}$, and \( (\text{S/N})_g > 30 \). With the selected sample, we first subtract the trend of radial gradient, and then show the vertical gradient of [Fe/H] across different $R$ ranges in Figure~\ref{fig:7}. The gradient ranges from $-0.079 \pm 0.077$ to $0.025 \pm 0.080 \, \text{dex} \, \text{kpc}^{-1}$, consistent with the previous analyses of the OBA stars and OCs, supporting  the conclusion that the vertical gradient for mono-age stellar populations is close to zero.

\section{Summary and Discussion}
We investigate the vertical gradient of [Fe/H] in the Galactic disk using data from the LAMOST-LRS young stellar sample. By analyzing mono-temperature stellar populations within this sample, we find that as stellar $T_{\mathrm{eff}}$ (age) increases (decreases), the vertical gradient of [Fe/H] approaches zero across different $R$ ranges. This trend is particularly evident in mono-age stellar populations and is further validated by the OCs and DCEPs. In contrast, older populations, such as G dwarfs, exhibit a more pronounced vertical gradient of [Fe/H]. The results align with the predictions of numerical simulations by \cite{Graf2024}, supporting the ``upside-down'' settling process of the Galactic disk over time.

The flat vertical metallicity gradient observed in mono-age stellar populations is primarily due to the regulation of the star-forming gas disk's vertical extent by turbulence. Turbulence not only effectively supports the thickness of the gas disk but also facilitates efficient mixing of metals, resulting in a uniform distribution of metallicity in the vertical direction. Consequently, mono-age stellar populations exhibit minimal variations in metallicity with vertical height, characterized by a nearly flat vertical metallicity gradient.

Similarly, when considering stars of all ages, the observed strong vertical metallicity gradient primarily reflects the ``upside-down'' settling process of the Galactic disk over time, accompanied by continuous metal enrichment. These vertical gradients progressively weaken with increasing radius, driven by the narrowing age range of stars, which in turn leads to a more constrained distribution of metallicity.

Compared to \cite{Xiang2015}, we adopt the mono-temperature stellar population method to define the vertical metallicity gradient of mono-age stellar populations more rigorously, and extend the relationship between the gradient and stellar age to young populations with $\tau < 1.5 \, \text{Gyr}$. This more precise age classification effectively reduces biases introduced by age mixing and accounts for the impact of radial gradients, allowing us to reveal the characteristics of vertical metallicity gradient with greater accuracy. Our study primarily focuses on stellar populations within the Galactocentric distance range $7 < R \leq 13$\,\text{kpc}, encompassing the transition region from the solar neighborhood to the outer disk. While this radial coverage captures key chemical evolution signatures predicted by the two-infall model beyond the solar circle \citep{Chiappini1997, Lian2020}, it is important to note that our current observations do not extend across the entire Galactic disk. Recent kinematic studies based on LAMOST data suggest that the disk structure of the MW may extend out to 35 kpc \citep{Tian2024}. Despite the spatial limitations of our study, it provides new insights into the chemical evolution of the MW and serves as a valuable complement to existing observational constraints.

\section*{Acknowledgments}
This work is supported by the National Natural Science Foundation of China through the projects NSFC 12222301 and 12173007, as well as by the National Key Basic R\&D Program of China (2024YFA1611901 and 2024YFA1611601). 

This work has made use of data from the European Space Agency (ESA) mission {\it Gaia} (\url{https://www.cosmos.esa.int/gaia}), processed by the Gaia Data Processing and Analysis Consortium (DPAC, \url{https://www.cosmos.esa.int/web/gaia/dpac/consortium}). The Guoshoujing Telescope (the Large Sky Area Multi-Object Fiber Spectroscopic Telescope, LAMOST) is a National Major Scientific Project built by the Chinese Academy of Sciences, with funding provided by the National Development and Reform Commission. LAMOST is operated and managed by the National Astronomical Observatories, Chinese Academy of Sciences.

\bibliographystyle{aasjournal}  
\bibliography{Article}  
\end{document}